Project Success in Agile Development Projects

Student: Alberto Perez Veiga

University of Maryland University College

IT Project Management (ITEC 640 - 9041)

November 2017

**Abstract**

The paper explains and clarifies the differences between Waterfall and Agile development methodologies, stablishes what criteria could be taken into account to properly define project success within the scope of software development projects, and finally tries to clarify if project success is the reason why many organizations are moving to Agile methodologies from other ones such as Waterfall. In the form of a literature review, it analyses several, publications, investigations and case studies that point out the motives why companies moved to Agile, as well as the results they observed afterwards. It also analyses overall statistics of project outcomes after companies evolved from traditional methodologies such as Waterfall to Agile development approaches.

**Research question**

Is project success the reason that the Agile methodology is replacing the Waterfall methodology in many organizations?

**Introduction**

Agile methodologies have driven many of the Software Development Lifecycles since the appearance of the Agile Manifesto. A wide range of companies implement different variations of Agile in order to more effectively develop software and meet customer's expectations faster and better than with the traditional waterfall approach.

When evaluating the outcome of a project, many criteria and metrics can be used. Although the typical success criteria would assume finishing the project within budget, time and scope, others such as client satisfaction should be also take into account. A project that is finished in time and budget may finish with an unsatisfied client. In other occasions, a project which spent over the estimated budget and over time may end up with a satisfied customer who considers the project a total success.

## Waterfall model

The process that is considered as the origin of the waterfall process or methodology was described by Dr. W. Royce in his work "Managing the development of large software systems", although he never used the word "waterfall", his process set the grounds for this methodology. This process is comprised of seven clear steps which, once completed in a pre-defined order, should result in a stable software product. These steps are: systems requirements, software requirements, analysis, design, coding, testing and operations (Royce, 1970).

The first phases of the process, before coding, usually take a long time, and requirements must be clearly defined before starting coding. Each of the steps mentioned by Dr. Royce has to be executed sequentially, meaning that each of them must be complete before proceeding to the next one. This represents what has been stated as one the big inconveniences in this model: the difficulty to adapt to changes in requirements or the lack of initial clarity on them, what leads to the need of constant work and rework during the life of the project (Verner & Cerpa, 1997). The lack of flexibility of this, the traditional model, is considered as one of its big hindrances.

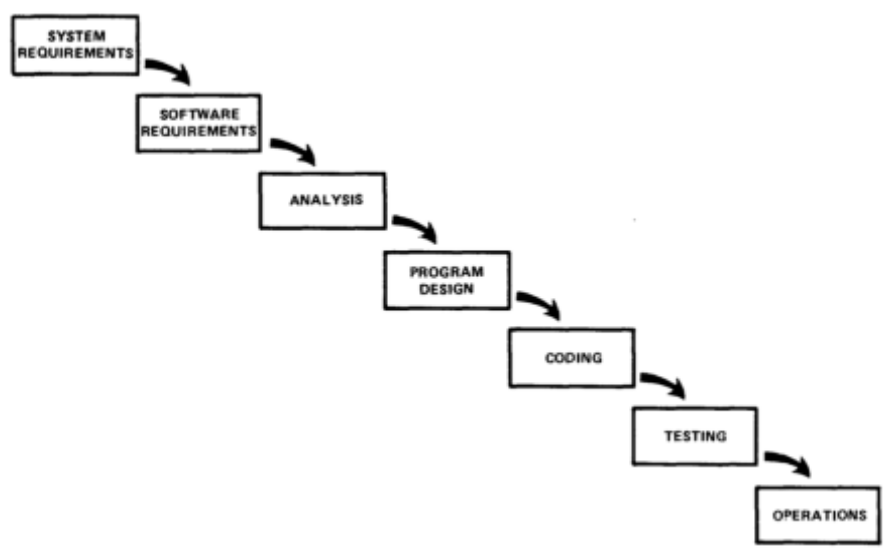

*Figure 1.* The seven steps of Waterfall (Royce, 1970)

## Agile model

The Agile concept was introduced by a group of developers in 2001 and summarized in The Agile Manifesto, which has the following principles as main cornerstones: "Individuals and interactions over processes and tools, working software over comprehensive documentation, customer collaboration over contract negotiation, and responding to change over following a plan" (Beck, 2001).

The Agile development methodology focuses on allowing programmers to develop quickly in the presence of uncertain or changing requirements (Martin, 2003). In general, the requirements are gathered and analyzed before the start of the coding work, however, they are revisited during each iteration. At predefined periods of time (i.e. less than one month in SCRUM) (Schwaber & Sutherland, 2016), requirements are revised through a close collaboration between the project team and relevant stakeholders such as project sponsor. The customer constantly validates the outcome of each iteration and provides refinements of the initial requirements.

In general, developing according to any of the Agile methodologies involves a strong focus on the business needs, strong collaboration principles, a compromise with quality, and the delivery of iterative developments which are fully functional and are ultimately presented to the different business roles, including the customer, in order to be refined over time. (Craddock, 2015). The process of achieving a final software product following an Agile methodology would be similar to Figure 2.

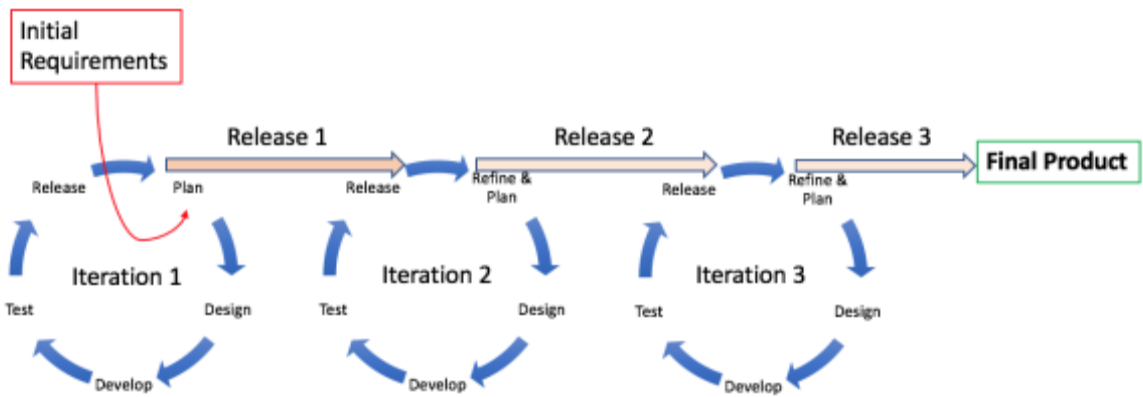

*Figure 2* Agile development process

Agile prescribes working collaboratively, in iterations and with flexibility. It doesn't provide, however, any specific framework. Some of the most well-known and usual Agile frameworks or methodologies are SCRUM, KANBAN and Extreme Programming. Although all of them share the Agile cornerstone values and principles, each of them has their own particularities.

Although the concept of Agile was introduced by in 2001, it wasn't until approximately 2008 when companies started widely adopting Agile methodologies. There was a sharp increase in its adoption between 2009 and 2015. The majority of the organizations revealed in 2015 that they were using Agile while only around 9% of them were still developing following a waterfall approach (Jeremiah, 2015).

**Success Criteria**

Traditionally, the three angles of the triple constraint, also known as the iron triangle, have been used to measure project success: time, budget and scope. A project is considered successful if it finishes in the planned time, using the planned and assigned scope and the expected features (scope) are delivered. According to the Standish Group Chaos Report, a study developed in 1994 over 365 collaborations and more than 8,000 applications, of large, medium and small companies, a project is considered challenged if it fails in one of the three constraints or impaired (failed) if it's completely canceled (Lech, 2013).

According to the 1994 Standish Group Chaos Report, the three main sources for project failure, as expressed by Information Technology (IT) executive managers were: Lack of User Input (12.8%), Incomplete Requirements and Specifications (12.3%), and Changing Requirements and Specifications (11.8%) (The Standish Group, 1994).

Literature suggests that success depends on multiple dimensions, although it's not clear which of the dimensions best represents or weights the most towards the definition of success (Joosten, Basten and Mellis, 2011). The use of the iron triangle to measure projects success is probably due to the fact that other factors such as customer satisfaction are not enough specific or measurable to constitute a rigorous comparable indication (Cuellar, 2010). Baccarini (1999) proposed that there should two different categories to measure success: one for the product success, which would entail measuring how the delivered product fulfils the customer's expectations in terms of organizational goals and objectives; and another one for project success, which would measure the outcomes of the project in terms of time, budget and features constraints.

Therefore, summarizing the statements above, project success as a whole shall meet two different criteria: *product success* and strictly *project success*. The first one represents how the project meets the customer's organizational or business goals and the latter represents an absolute measurement on how the project matches the traditional success criteria as expressed in the triple constraint.

## Product Success

Product success shall be measured, as mentioned, in terms of how the organizational needs of the customer have been fulfilled. I this respect, taking as an example, the enterprise implementation of an Enterprise Resource Planning software (ERP), holding SCRUM meetings on a timely basis with users of the system (the internal customers), makes the development of requirements faster, and more accurate, as opposed to the traditional requirements gathering at the beginning of the project. Additionally, by keeping releases short, around two weeks, increases the chances of finding errors early in the development cycle (Doig, 2015). Customers are involved during the whole lifecycle; thus, the chances of customer satisfaction are higher.

Some of the reasons considered very important or of the highest importance by companies adopting Agile are, among others: Accelerate time to market, manage changing priorities, better align IT with business, and enhance software quality (Vision One, 2014), which are some of the criteria that are linked to the main goal of product success: to better contribute to business goals and objectives. Additionally, a study developed by HP in 2014 over 600 developers and IT professionals show that one of the highest-ranking reasons (49%) for organizations to adopt agile, is that they believe that its adoption results in increased customer satisfaction (Jeremiah, 2017).

## Project Success

The 2015 Standish Group Chaos Report, which summarizes data from the outcomes of more than 10,000 projects between 2011 and 2015, highlights that, in general, for projects of all sizes, Agile projects are more successful than waterfall projects. This tendency is more noticeable for medium and large sized projects, rather than for smaller ones, where the success ratios seem to be closer (Hastie, 2015). Table 1shows the figures of the study.

| Size | Method | Successful | Challenged | Failed |
|------|--------|------------|------------|--------|
| All sizes | Agile | 39 | 52 | 9 |
| | Waterfall | 11 | 60 | 29 |
| Large | Agile | 18 | 59 | 23 |
| | Waterfall | 3 | 55 | 42 |
| Medium | Agile | 27 | 62 | 11 |
| | Waterfall | 7 | 68 | 25 |
| Small | Agile | 58 | 38 | 4 |
| | Waterfall | 44 | 45 | 11 |

*Table 1* Resolution of software projects 2011-2015 (Hastie, 2015)

Kroger is one of the largest groceries companies in the USA, with sales over $90 Billion in the fiscal year 2011. Although apparently, the waterfall processes of the company were working correctly, especially for the projects that had well-defined constraints, most projects were delivered over cost, consuming more resources than initially planned and many of them failed to meet customer's expectations, not delivering the value expected from them. Although Kruger struggled with the transformation and went through different unsuccessful attempts, when they committed to the adoption of Agile, after the first wave of their movement from a pure waterfall, 3-tiered software development approach, they realized the following benefits:

- An acceleration of 18,5% in their development cycle, as a consequence of the better collaboration between the project owner and development teams. Kruger went from delivering iterations on 30 months average, to smaller software delivery in less than three months each. They also pointed out increased trust and greater business satisfaction as additional benefits.

- A steady expenditure of the budget between 12-13% in each iteration, what lead to increased trust from the customer, strictly controlled budget, times and scopes and consequently, additional levels of customer satisfaction. (Smith, 2012)

## Conclusion

The adoption of Agile methodologies, since their appearance has been on the rise across industry. The continuous implementations of methodologies such as SCRUM, KANBAN or Extreme Programming, is a stablished trend since 2008, which reached a point around 2015 where the majority of companies seem to either have fully adopted Agile, or use it for most of their development projects.

Project success is complex to measure. Trying to classify projects outcome attending to the traditional iron triangle criteria would be the most mathematical, concrete way to measure it, however companies seem to start associating project success to a few other criteria probably more complicated to evaluate, such as customer satisfaction or IT alignment with business goals and objectives.

Customer's requirements are almost never fixed and tend to change over time, either due to new business needs, technological advances or simply changes on the priorities of the company. This is particularly critical in larger projects. Organizations have understood that developing using Waterfall does not provide enough flexibility to meet those changing requirements according to stakeholder's expectations. Agile, in general advocates for flexible development, adapted to changing customer needs, increases software quality, and provides an overall tangible increase in customer satisfaction. These are some of the main reasons for organizations moving to Agile, according to different studies and surveys.

Sticking to the more traditional approach of measuring success in terms of budget, features and time, outcomes are consistently better for Agile developed software projects, especially for medium and large-size projects. Although for the smallest projects differences are not so big, they benefit from the adoption of Agile methodologies too. In general, literature suggests that not only Agile developments actually have a higher ratio of project success (intended as product and project success), but also organizations, in general, have understood this, and they adopt Agile as a way to achieve better ratios of project success.